\documentclass[prb,twocolumn,superscriptaddress,showpacs]{revtex4}
\pdfoutput=1
\usepackage{graphicx}
\usepackage[pdftex]{hyperref}
\usepackage{amsmath}
\usepackage{amssymb}
\usepackage{amsopn}
\usepackage{amsbsy}
\usepackage{braket}
\usepackage{mathrsfs}
\newcommand{\pprl}{Phys. Rev. Lett.\ }
\newcommand{\pprb}{Phys. Rev. B\ }

\newcommand{\p}{\partial}

\newcommand{\ud}{\,\text{d}}

\renewcommand{\v}[1]{\mathbf{#1}}

\newcommand{\zb}{\bar{z}}
\usepackage{slashed}
\usepackage{appendix}
\begin{document}
\title[Composite Dirac Fermions]{Dynamical Mass Generation of Composite Dirac Fermions and Fractional
  Quantum Hall Effects near Charge Neutrality in Graphene}
\author{Feng Cai}
\affiliation{Department of Physics, Boston College, Chestnut Hill,
  MA 02467}
\author{Yue Yu}
\affiliation{
  The State Key Laboratory of Theoretical Physics, Institute of Theoretical
  Physics, Chinese Academy of Sciences, P.O. Box 2735, Beijing 100190, China}
\author{Ziqiang Wang}
\affiliation{Department of Physics, Boston College, Chestnut Hill,
  MA 02467}
\date{\today}
\begin{abstract}
  We develop a composite Dirac fermion theory for the fractional quantum
  Hall effects (QHE) near charge neutrality in graphene. We show that the
  interactions between the composite Dirac fermions lead to a dynamical mass
  generation through exciton condensation. The four-fold spin-valley
  degeneracy is fully lifted due to the mass generation and the exchange
  effects such that the odd-denominator fractional QHE observed in the
  vicinity of charge neutrality can be understood in terms of the integer
  QHE of the composite Dirac fermions. At the filling factor $\nu=1/2$, we
  show that the massive composite Dirac fermion liquid is unstable against
  chiral $p$-wave pairing for weak Coulomb interactions and the ground state
  is a paired nonabelian quantum Hall state described by the Moore-Read
  Pfaffian in the long wavelength limit.
\end{abstract}
\maketitle

\section{Introduction}
\label{sec:introduction}

Since the discovery of graphene~\cite{graphene}, a rich set of integer
\cite{Novoselov,YZhang1,YZhang2,Jiang} and fractional
\cite{Du,Bolotin,Ghahari,Dean} quantum Hall effects (QHE) has been explored
experimentally under the new setting of two-dimensional lattice electrons
with a relativistic energy-momentum dispersion at low energy. Currently, the
$\nu=1/3$ state has been convincingly
observed\cite{Du,Bolotin,Ghahari,Dean,Feldman1,Feldman2}. Remarkably, initial evidence for the
emergence of a $\nu=1/2$ fractional quantum Hall state has been reported
\cite{Ghahari}, raising the hope for realizing nonabelian statistics in
suspended graphene.

For graphene, it is known that there is a spin-valley SU(4) symmetry. This
restricts the filling factors $\nu$ of the integer QHE of the Dirac fermions
to $\pm2,\pm6,\pm10,\cdots$, provided that the SU(4) symmetry is
unbroken. The observation of the $\nu= \pm 1$ integer Hall
plateaus~\cite{YZhang1} indicates that the SU(4) symmetry is
broken. Dynamical mass generation, which lifts the spin-valley degeneracy of
the zeroth Dirac Landau level ($\zeta LL$)~\cite{mc}, as well as quantum
Hall ferromagnetism~\cite{MacDonald,Alicea,KYang} have been proposed for its
explanation. Despite recent theoretical efforts on the fractional QHE (FQHE)
in graphene~\cite{KYang,Apalkov,Toke1,Goerbig,Khveshchenko,Papic}, it
remains uncertain whether the observed $\nu=1/3$ state implies SU(4)
symmetry breaking and is therefore a single-component Laughlin state, or a
multi-component Halperin state with spin-valley
degeneracy~\cite{Papic}. Moreover, although states at even-denominator
filling factors have been investigated numerically~\cite{Toke2,Wojs} by
exact-diagonalization, the effect of Landau level mixing, which may be
relevant for the stabilization of the Moore-Read Pfaffian
state~\cite{BN,WTJ,Rezayi,Sodemann,Peterson,Simon}, was neglected when the
Hilbert space is projected and restricted to that of a specific Landau
level.

In this paper, we propose a mechanism of dynamical mass generation by
exciton condensation and exchange-driven polarization to qualitatively
describe both the abelian and the nonabelian FQHE in graphene. To this end,
we extend the composite fermion Chern-Simons (CS) theory \cite{HLR,Jain} to
the case of Dirac particles attached to an even number of flux quanta
through the CS gauge field, which can be implemented by a unitary
transformation. We will refer to this transformation as the CS
transformation.

The CS approach is more suitable for studying the fractional quantum Hall
regime where Landau level mixing is significant, since it does not restrict
the quantum states to the $\zeta LL$. The importance of Landau level mixing
in graphene can be seen from the fact that, contrary to semiconductor
heterostructures where the dispersion is non-relativistic, the Coulomb
interaction energy and Landau level spacing for Dirac fermions both scale
with $\sqrt{B}$, where $B$ is the strength of the magnetic
field. Specifically, the ratio of the interaction energy and energy spacing
between the zeroth and first Landau levels is given by
$\alpha_g/\sqrt{2}$ where $\alpha_g = e^2/4\pi\epsilon \hbar v_F$ is the
fine structure constant for graphene. For free standing graphene, the bare
value of the fine structure constant is $\alpha_g\approx 2.2$ (see e.g.,
Ref.~\cite{Drut}). Therefore, the two energy scales are of the same order;
the system should be in the strong coupling regime~\cite{mc} where Landau
level spacing is a relevant energy scale.

A crucial step in our theory is to work with the proper particle density via
a particle-hole transformation such that the vacuum state of the
relativistic composite Dirac fermions (CDF) is defined by the charge neutral
state with all negative energy states filled. The CDF theory introduced in
this paper are rather general, involving only relations between the CDF
particle density and its filling fraction $\widetilde{\nu}$. As we will
show, the latter corresponds to a unique filling fraction $\nu$ of the
original electrons once the ground state is determined.

At the filling fractions $\widetilde\nu= \pm 1/\widetilde\phi$ with
$\widetilde\phi$ an even integer, we will show, by variational calculations
of the ground state energy, that it is energetically favorable for the CDF
to develop an exciton condensate. The latter supports single quasiparticle
excitations with a CDF mass gap at low density above the condensate. It
turns out that the possible CS transformations can be grouped into three
types depending on whether the SU(4) symmetry is broken and how it is
broken. If the exciton mass gaps have the same sign for all SU(4)
components, the SU(4) symmetry is preserved. This kind of CDF systems have a
non-zero Chern number $\pm2$, implying that the exciton condensate is an
anomalous Hall liquid. In terms of the Bloch electrons in graphene, this
corresponds to a total filling factor $\nu=\pm(2-\widetilde\nu)$. However,
when the exciton mass gaps for different SU(4) components are of different
signs, the SU(4) symmetry is broken and the spin-valley degeneracy is
lifted. There are two ways to break the symmetry since we have two
inequivalent ways to partition the four spin-valley components into two
groups. In the first case, one of the components has a different sign from
the remaining three; in the second case, the four components are divided
evenly into two groups. The former has Chern number $\pm 1$ and describes
electron states at $\nu = \pm (1 - \widetilde\nu)$. The latter is
topologically trivial with net Chern number zero. It therefore describes the
electron states at filling fractions $\nu = \pm \widetilde \nu$. In this
case, the exchange term in the statistical interaction further breaks the
remaining symmetry so that the ground state is fully spin-valley polarized.

Thus, in the present theory, the dynamical SU(4) symmetry-breaking mass
generation offers a route to the observed FQHE states in graphene near
charge neutrality~\cite{Du,Bolotin,Ghahari,Dean}. We show that the
quasiparticles above the exciton condensate have a pairing instability in
the chiral $p$-wave channel, leading to an even denominator paired quantum
Hall state described by the nonabelian Moore-Read Pffafian~\cite{MR,RG} in
the long wavelength limit. On the other hand, at odd-denominator filling
fractions such as $\nu=1/3$, the quasiparticles above the exciton condensate
occupy fully filled Landau levels of the residual magnetic field
\cite{Jain}, giving rise to a single-component Laughlin state for the
electrons in graphene.

The key ingredient of the present theoretical framework is the dynamical
mass generation of massless Dirac fermions, which enables us to approach the
FQHEs near filling factors $\nu = 0, \pm 1, \pm 2$ in a unified way. This is
an exciting example of the interplay between condensed matter and high
energy physics; as the mechanism was introduced by Nambu and
Jona-Lasinio~\cite{Nambu} inspired by the BCS theory of
superconductivity. It was used recently to explain Landau level
splitting~\cite{mc} in graphene. Here we extend this mechanism to the FQHE
of lattice electrons in graphene, thus provide another physical realization
of dynamical mass generation in a condensed matter system. We note that such
an excitonic mass generation intrinsically involves Landau level mixing
since it requires the formation of particle-hole pairs consisting of states
from different Landau levels.

This paper is structured as follows. In Section \ref{sec:cdf}, we formulate
the composite Dirac fermion theory by introducing a unitary transformation
that implements the flux attachment. We then derive the statistical
interaction mediated by the CS gauge field, focusing on the even denominator
filling factors where the external magnetic field is canceled by the flux of
the CS field. The effects of the statistical interaction is then
investigated. In Section \ref{sec:exciton}, we show that an immediate
consequence of the statistical interaction is the exciton condensation which
opens up mass gaps for the four components of the CDFs. The normal state is
obtained by doping the exciton insulating state according to the filling
factors of the fractional quantum Hall states. In Section \ref{sec:ps}, we
study the pairing instability induced by the statistical interaction. We
show that the leading instability is in the complex $p$-wave channel which
provides a realization of the nonabelian Moore-Read state in graphene.

\section{Theory of composite Dirac fermions}
\label{sec:cdf}

We start with graphene electrons in the continuum limit under a
perpendicular external magnetic field $\v B = -B\hat{\v z}$. The effective
Hamiltonian can be written down in terms of the four-component fermion
operator $\psi_s(\v x)$ for the two-sublattice and two-valley degrees of
freedom,
\begin{align}
  \label{eq:Hamiltonian}
  H =
  -i\hbar v_F\int\ud^2x\psi_s^\dagger\alpha_i(\p_i +
  ieA_i/\hbar)\psi_s,
\end{align}
where $v_F$ is the Fermi velocity and $A_i(\v x)$ is the vector potential:
$\nabla\times \v A=\v B$. The summations over repeated spin index $s$ and
the spatial index $i=1,2$ are implied. The $4\times4$ matrix $\alpha_i =
\gamma^0\gamma^i$, with the $\gamma$ matrices given by $\gamma^0 = I\otimes
\tau_1$, $\gamma^i = -i\sigma_i\otimes\tau_2$, where $\sigma_i$ and $\tau_i$
are the $2\times2$ Pauli matrices acting in the sublattice and valley
subspaces, respectively. To make the SU(4) symmetry explicit, it is
instructive to separate out the valley degrees of freedom and rewrite the
Hamiltonian as
\begin{align}
  \label{eq:4-hamiltonian}
  H &= -i\hbar v_F \int\ud^2x\psi_{Rs}^\dagger(\v x)\sigma_i\left(\p_i +
    i\frac{e}{\hbar}A_i\right)\psi_{Rs}(\v x)\nonumber\\
  &+ i\hbar v_F\int\ud^2x\psi_{Ls}^\dagger(\v x)\sigma_i\left(\p_i +
    i\frac{e}{\hbar}A_i\right)\psi_{Ls}(\v x).
\end{align}
Here $\psi_{\tau s}(\v x)$ is a two-component fermion spinor field for spin
$s=\uparrow,\downarrow$ and valley $\tau=R,L$. For convenience, we denote
$\psi_{\tau s}=\psi_{\alpha}$ with $\alpha=\{1, 2, 3, 4\} = \{(\downarrow,
R), (\downarrow, L), (\uparrow, R), (\uparrow, L)\}$.

The CDF field $\Psi_\alpha$ can be introduced by a unitary transformation
\begin{align}
  \label{eq:unitary_transformation}
  \left(\begin{array}{c}
      \psi_1 \\
      \psi_2\\
      \psi_3\\
      \psi_4
    \end{array}\right)
  =
  \begin{pmatrix}
    e^{i\mathcal{I}_1} && 0 && 0 && 0\\
    0 && e^{i\mathcal{I}_2} && 0 && 0\\
    0 && 0 && e^{i\mathcal{I}_3} && 0\\
    0 && 0 && 0 && e^{i\mathcal{I}_4}\\
  \end{pmatrix}
  \left(\begin{array}{c}
      \Psi_1 \\
      \Psi_2\\
      \Psi_3\\
      \Psi_4
    \end{array}\right),
\end{align}
with
\begin{align}
  \label{eq:I}
  {\cal I}_\alpha(\v x) =
  \sum_{\beta}\int\ud^2 x'\mathcal{K}_{\alpha\beta}\rho_{\beta}(\v x')\arg(\v x - \v x'),
\end{align}
where $\alpha,\beta=1,2,3,4$ and $\rho_\alpha$ is the particle density
operator in the spin-valley sector $\alpha$. Note that due to the presence
of multiple components, a $\mathcal{K}$-matrix must be introduced in the
unitary transformation~\cite{Wen}. Its physical meaning is explained
below. The form of the $\mathcal{K}$-matrix can be specified by physical
considerations. In order for the CDF field $\Psi_\alpha(\v x)$ to be
fermionic, $\mathcal K$ must be symmetric and its diagonal elements must be
even~\cite{Rajaraman}.

The transformed Hamiltonian, i.e. the Hamiltonian of the CDFs is given by
\begin{align}
  \label{eq:hamiltonian_CDF}
  H = (-1)^\alpha i\hbar v_F\int\ud^2x\Psi_\alpha^\dagger\sigma_i\left[\p_i + i{e\over\hbar} (A_i +
    a^\alpha_i)\right]\Psi_\alpha,
\end{align}
with the CS gauge field $\v a^\alpha=(a_1^\alpha, a_2^\alpha)$:
\begin{align}
  \label{eq:statistical_gauge_field}
  a^\alpha_1(\v x) &= \frac{\hbar}{e}\sum_{\beta}\mathcal{K}_{\alpha\beta}\int\ud^2x'
  \frac{-(x_2 - x'_2)}{|\v x - \v x'|^2}\rho_{\beta}(\v x'),\nonumber \\
  a^\alpha_2(\v x) &= \frac{\hbar}{e}\sum_{\beta}\mathcal{K}_{\alpha\beta}\int\ud^2x'
  \frac{x_1 - x'_1}{|\v x - \v x'|^2}\rho_{\beta}(\v x').
\end{align}
It is straightforward to verify that $\v a^\alpha$ satisfies
\begin{align}
  \label{eq:cs}
  \nabla\times\v a^\alpha(\v x) = \hat{\v z}\sum_{\beta}\mathcal{K}_{\alpha\beta}\rho_{\beta}(\v
  x)h/e.
\end{align}
Physically, Eq.~\eqref{eq:cs} describes how the CS gauge field coupled to
the CDFs with the spin-valley index $\alpha$ is generated by the flux quanta
attached to other CDFs in the same or different spin-valley sectors. The
number of flux quantum is specified by the matrix elements of
$\mathcal{K}$. We note that $\rho_\alpha$ can be either positive or negative
due to the particle-hole symmetry of the Dirac spectrum and our choice of
vacuum associated with the charge neutral point. As a result, the
matrix elements of $\mathcal K$ can be positive or negative. In particular,
if $\mathcal K$ describes a state with filling factor $\widetilde{\nu}$,
$-\mathcal K$ describes a state with filling factor $-\widetilde \nu$. The
condition for the CDFs to experience a vanishing net magnetic field on
average requires
\begin{align}
  \label{eq:cancellation}
  \sum_\beta \mathcal{K}_{\alpha\beta}\braket{\rho_\beta} = eB/h
\end{align}
for all the values of $\alpha$. For a given filling fraction $\widetilde\nu
= \sum_\alpha\braket{\rho_\alpha}h/eB = \pm
1/\widetilde{\phi}$, there are three types of physical solutions:
\begin{align}
  \label{eq:kmatrix}
  \mathcal{K}^1 &= \mathrm{sgn}(\widetilde\nu)
  \begin{pmatrix}
    \widetilde{\phi} & \widetilde{\phi} & \widetilde{\phi} &
    \widetilde{\phi}\\
    \widetilde{\phi} & \widetilde{\phi} & \widetilde{\phi} &
    \widetilde{\phi} \\
    \widetilde{\phi} & \widetilde{\phi} & -\widetilde{\phi} &
    -\widetilde{\phi} \\
    \widetilde{\phi} & \widetilde{\phi} & -\widetilde{\phi} &
    -\widetilde{\phi}
  \end{pmatrix}, \nonumber \\
  \mathcal{K}^2 &= \mathrm{sgn}(\widetilde\nu)
  \begin{pmatrix}
    \widetilde{\phi} & \widetilde{\phi} & \widetilde{\phi} &
    \widetilde{\phi}\\
    \widetilde{\phi} & \widetilde{\phi} & \widetilde{\phi} &
    \widetilde{\phi} \\
    \widetilde{\phi} & \widetilde{\phi} & \widetilde{\phi} &
    \widetilde{\phi} \\
    \widetilde{\phi} & \widetilde{\phi} & \widetilde{\phi} &
    -\widetilde{\phi}
  \end{pmatrix},\nonumber \\
  \mathcal{K}^3 &= \mathrm{sgn}(\widetilde\nu)
  \begin{pmatrix}
    \widetilde{\phi} & \widetilde{\phi} & \widetilde{\phi} &
    \widetilde{\phi}\\
    \widetilde{\phi} & \widetilde{\phi} & \widetilde{\phi} &
    \widetilde{\phi} \\
    \widetilde{\phi} & \widetilde{\phi} & \widetilde{\phi} &
    \widetilde{\phi} \\
    \widetilde{\phi} & \widetilde{\phi} & \widetilde{\phi} &
    \widetilde{\phi}
  \end{pmatrix}.
\end{align}

An obvious distinction among the $\mathcal K$ matrices is that
$\mathcal{K}^1$ and $\mathcal{K}^2$ break the $\mathrm{SU}(4)$ symmetry
while $\mathcal{K}^3$ preserves the full symmetry. This can be seen from the
CS transformation~\eqref{eq:unitary_transformation}. If all four spin-valley
components rotate in the same way, $\mathrm{SU}(4)$ symmetry is
preserved. This requires the operator $\mathcal I_\alpha(\v x)$ to be
identical for the four components. This condition, together with the general
requirements for the $\mathcal K$-matrix mentioned above, uniquely
determines $\mathcal K^3$. On the other hand, the CS transformation
associated with $\mathcal K^1$ rotates the two spin components differently,
thus leads to a broken $\mathrm{SU}(2)$ spin symmetry. Furthermore, the
solution of Eq.~\eqref{eq:cancellation} for $\mathcal{K}^1$ requires
$\braket{\rho_{\downarrow}} = eB/h\widetilde{\phi}$,
$\braket{\rho_{\uparrow}} = 0$ for $\widetilde\nu = 1/\widetilde{\phi}$,
where $\rho_\downarrow = \rho_1 + \rho_2$ and $\rho_\uparrow = \rho_3 +
\rho_4$. Thus $\mathcal{K}^1$ describes a state in which spin is fully
polarized. Equivalently, one can interchange the spin and valley to break
the valley symmetry this way, resulting in a $\mathcal{K}$-matrix of similar
structure as $\mathcal{K}^1$. Interestingly, the $\mathrm{SU}(4)$ symmetry
can be broken in another way, i.e., by keeping any three of the spin-valley
components degenerate while rotating the remaining component in a different
way. This is achieved by the $\mathcal K^2$. Similar to $\mathcal K^1$, this
matrix implies a constraint $\braket{\rho_4} = 0$ when $\widetilde{\nu}
=1/\widetilde{\phi}$. We will show that the CDFs have different topological
properties when they are in the states specified by these three
characteristic $\mathrm{K}$-matrices. Note that, due to the particle-hole
symmetry, the $\mathcal{K}$-matrices for positive (particle) and negative
(hole) filling factors have opposite signs. As a result, some of the
$\mathcal{K}$-matrices have negative eigenvalues. This should be contrasted
to the case of non-relativistic multi-component electron systems where the
$\mathcal{K}$-matrices with negative eigenvalues result in wavefunctions
that are not normalizable~\cite{Beugeling}, as a consequence of the
quadratic energy-momentum dispersion.

The expressions for $\mathcal K^1$ and $\mathcal K^3$ can be further
simplified by noting that the two valleys for a given spin projection are
degenerate in these states. As a result, the $\mathcal{K}$-matrices in this
case effectively reduces to $2\times 2$ matrices. Let $\Psi_\downarrow =
(\Psi_1,\Psi_2)^T$ and $\Psi_\uparrow = (\Psi_3,\Psi_4)^T$, the CS
transformation can be expressed as
\begin{align}
  \label{eq:unitary_transformation_2x2}
  \left(\begin{array}{c}
      \psi_\downarrow(\v x) \\
      \psi_\uparrow(\v x)
    \end{array}\right)
  =
  \begin{pmatrix}
    e^{i\mathcal{I}_\downarrow(\v x)} && 0\\
    0 && e^{i\mathcal{I}_\uparrow(\v x)}
  \end{pmatrix}
  \left(\begin{array}{c}
      \Psi_\downarrow(\v x) \\
      \Psi_\uparrow(\v x)
    \end{array}\right),
\end{align}
with
\begin{align}
  \label{eq:I_2x2}
  {\cal I}_s(\v x) =
  \sum_{s'}\int\ud^2 x'\mathcal{K}_{ss'}\rho_{s'}(\v x')\arg(\v x - \v x'),
\end{align}
and
\begin{equation}
  \label{eq:kmatrix_2x2}
  \mathcal{K}^1 =
  \begin{pmatrix}
    \widetilde{\phi} & \mathrm{sgn}(\widetilde\nu)\widetilde{\phi}\\
    \mathrm{sgn}(\widetilde\nu)\widetilde{\phi} & -\widetilde{\phi}
  \end{pmatrix}, \quad
  \mathcal{K}^3 =\mathrm{sgn}(\widetilde\nu)
  \begin{pmatrix}
    \widetilde{\phi} & \widetilde{\phi}\\
    \widetilde{\phi} & \widetilde{\phi}
  \end{pmatrix}.
\end{equation}
Under the transformation,
\begin{align}
  \label{eq:hamiltonian_CDF_2x2}
  H = -i\hbar v_F\int\ud^2x\Psi_s^\dagger\alpha_i(\p_i + ie (A_i +
  a^s_i)/\hbar)\Psi_s.
\end{align}
To keep the presentation simple, we will focus on $\mathcal{K}^1$ and
$\mathcal{K}^3$, and comment on $\mathcal{K}^2$ when appropriate.

The Hamiltonian $H$ in Eq.~\eqref{eq:hamiltonian_CDF_2x2} can be separated into two parts
\begin{align}
  H = H_0 + V_{\mathrm{st}},
\end{align}
where $H_0$ is the Hamiltonian for free massless Dirac particles and
$V_{\mathrm{st}}$ describes the CDF interactions mediated by the CS gauge field. It can
be obtained from the $A_i + a^s_i$ terms by using the explicit expressions for
$a^s_i$ in Eq.~\eqref{eq:statistical_gauge_field} and writing the
vector potential $A_i$ in terms of the average density
$\braket{\rho_s}$. The final result is an interaction of the form:
\begin{align}
  \label{eq:statistical_interaction}
  V_{\mathrm{st}}
  =& \sum_s-i\hbar v_F\int \ud^2x\ud^2x'\left(\Psi^\dagger_{Rs}(\v
    x)\mathcal{M}_s(\v x,\v x')\Psi_{Rs}(\v x)\right.\nonumber\\
  &\left. - \Psi^\dagger_{Ls}(\v x)\mathcal{M}_s(\v x,\v x')\Psi_{Ls}(\v
    x)\right),
\end{align}
where,
\begin{align*}
  \mathcal{M}_s(\v x, \v x') =\sum_{s'}\mathcal{K}_{ss'}\delta\rho_{s'}(\v x')
  \begin{pmatrix}
    0 & 1/(z - z')\\
    -1/(\zb - \zb') &0
  \end{pmatrix},
\end{align*}
with the bilinear fermion operator
$\delta\rho_s=\rho_s-\langle\rho_s\rangle$, and the holomorphic coordinates
$z=x+iy$, $\zb=x-iy$. We will refer to $V_{\mathrm{st}}$, which is
essentially a CDF current-density interaction, as the statistical
interaction.

Since the CDF experiences zero net magnetic field, its
field operator can be conveniently expanded in the helicity basis as
\begin{align*}
  \Psi_{Rs}(\v x) &=
  \frac{1}{\sqrt{2\mathcal{V}}}\sum_{\v k}e^{i\v k\cdot\v x}\left[
    \begin{pmatrix}
      e^{-i\theta_{\v k}}\\
      1
    \end{pmatrix}
    A_{s\v k} +
    \begin{pmatrix}
      e^{-i\theta_{\v k}}\\
      -1
    \end{pmatrix}
    B^\dagger_{s\v k}\right],\\
  \Psi_{Ls}(\v x) &=
  \frac{1}{\sqrt{2\mathcal{V}}}\sum_{\v k}e^{i\v k\cdot\v x}\left[
    \begin{pmatrix}
      e^{-i\theta_{\v k}}\\
      -1
    \end{pmatrix}
    C_{s\v k} +
    \begin{pmatrix}
      e^{-i\theta_{\v k}}\\
      1
    \end{pmatrix}
    D^\dagger_{s\v k}\right],
\end{align*}
where $\theta_{\v k}= \arctan k_y/k_x$. Note that, in the operator
expansion, we have performed a particle-hole transformation such that
removing a particle in a negative energy state is redefined as creating a
hole with positive energy. Specifically, $B^\dagger_{s\v k}$ and
$D^\dagger_{s\v k}$ in the above expressions are the hole creation operators
for the $R$- and $L$-valleys respectively. In the helicity basis, the
kinetic energy part of the Hamiltonian becomes
\begin{align}
  H_0 = \sum_{s\v k}\hbar v_F
  k(A^\dagger_{s\v k}A_{s\v k} + B^\dagger_{s\v k}B_{s\v k} + C^\dagger_{s\v
    k}C_{s\v k} + D^\dagger_{s\v k}D_{s\v k}).
\end{align}

\section{Exciton condensation and normal states}
\label{sec:exciton}

The statistical interaction derived in the previous section has important
physical consequences on the nature of the ground states for the CDFs. We
now show that it drives the formation of an exciton condensate of the
CDFs. The normal state is obtained by populating quasi-particles on top of
the exciton condensate. In other words, the normal state of the CDFs
corresponds to doping an excitonic insulator. We shall use the variational
approach and construct a variational wavefunction for the normal. First, at
charge neutrality, the exciton vacuum can be written as follows,
\begin{align}
  \label{eq:exciton_vacuum}
  \ket{0} =& \prod_{s}\prod_{\v k}\left(\cos \varphi_{Rs\v k }
    - \sin \varphi_{Rs\v k}A^\dagger_{s\v k}B^\dagger_{s\v k}\right)\nonumber \\
  &\times\left(\cos \varphi_{Ls\v k}
    - \sin \varphi_{Ls\v k}C^\dagger_{s\v k}D^\dagger_{s\v k}\right)\ket{\mathrm{vac}},
\end{align}
where $\varphi_{R,L}$ are variational parameters and $\ket{\mathrm{vac}}$
describes the state where the valence CDF bands are filled and the
conduction bands empty. The quasiparticle operators associated with the
exciton condensate $|0\rangle$ can be obtained through a Bogoliubov
transformation and are given by,
\begin{align}
  \label{eq:exciton_excitations}
  a^\dagger_{s\v k} &= \cos\varphi_{Rs\v k}A^\dagger_{s\v k} + \sin\varphi_{Rs\v
    k}B_{s\v k},\nonumber \\
  b_{s\v k} &= -\sin\varphi_{Rs\v k}A^\dagger_{s\v k} +
  \cos\varphi_{Rs\v k}B_{s\v k},
\end{align}
for the $R$-valley and similarly in terms of $c^\dag_{s\v k}$ and $d_{s\v
  k}$ for the $L$-valley (not shown). It is straightforward to verify that
$\ket{0}$ contains no quasiparticles, i.e. $a_{s\v k}\ket{0} = b_{s\v
  k}\ket{0} = 0$.

The normal state with a nonzero particle density can be constructed by
creating quasiparticles on top of the exciton vacuum $\ket{0}$. A generic
normal state with a positive filling factor can be written as $\ket{N} =
\prod_s\prod_{k \le k^{Rs}_F}a^\dagger_{s\v k}\prod_{k\le
  k^{Ls}_F}c^\dagger_{s\v k}\ket{0}$ where $k_F^{Rs}$ and $k_F^{Ls}$ are the
Fermi wave vectors associated with the two valleys with spin $s$. At the
filling factor $\widetilde \nu$, we have
\begin{align}
  \label{eq:density}
  \sum_{s,k \le k^{Rs}_F}\braket{N|a^\dagger_{s\v k}a_{s\v k}|N} + \sum_{s,k \le
      k^{Ls}_F}\braket{N|c^\dagger_{s\v k}c_{s\v k}|N} = \widetilde{\nu}eB/h.
\end{align}
Hereafter, the wave vector will be measured in unit of $k_F =
1/l_B\widetilde{\phi}^{1/2}$, which corresponds to the spin-polarized CDF
Fermi vector, and the energy in unit of $\hbar v_F k_F$. We will show that
the variational energy is indeed minimized at nonzero $\varphi_{R,L}$ in
favor of an exciton condensate.

\begin{figure}[htb]
  \centering
  \includegraphics[width=0.47\textwidth]{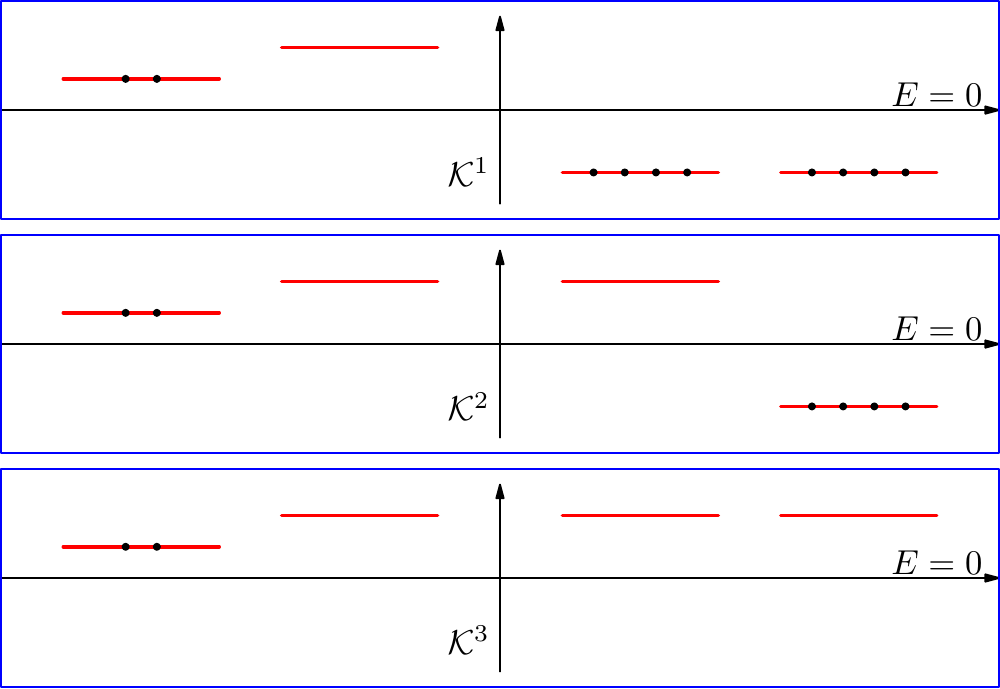}
  \caption{(color online) Schematic plots of the zeroth Landau levels for
    different $\mathcal{K}$-matrices. The lines with four dots depict
    completely filled Landau levels, and lines with two dots means partially
    filled Landau levels. The difference between the $\mathcal{K}$ matrices
    is reflected in the signs of the exciton mass gaps, and thus the
    relative positions of the zeroth Landau levels with respect to zero
    energy. $\mathcal{K}^1$ leads to opposite masses for the two spin
    projections and therefore describes the $\nu = 1/\widetilde{\phi}$ state
    (top panel). $\mathcal{K}^2$ has one component with a different sign,
    corresponding to the filling factor $\nu = -1 + 1/\widetilde{\phi}$
    (middle panel). $\mathcal{K}^3$ results in the same sign of mass for all
    bands, leading to the filling factor $\nu = -2 + 1/\widetilde{\phi}$
    (bottom panel).}
  \label{fig:landaulevels}
\end{figure}

It is useful to note that the exciton condensation energy from the
statistical interaction is linear in the exciton mass. This is due to the
fact that $V_{\mathrm{st}}$ is a current-density interaction and can be
expressed as,
\begin{align}
  \label{eq:stmass}
  &  V_{\mathrm{st}}
  = i\sum_{s,s'}\mathcal{K}_{ss'}\nonumber \\
  &\times \int\ud^2x\ud^2x'\bar{\Psi}_s(\v
  x)\gamma^3\gamma^5\Psi_{s'}(\v x')\frac{(\v x - \v x')\cdot\mathbf{J}_{s's}(\v x',\v x)}{|\v x - \v
    x'|^2},
\end{align}
where $J_{i,s's}(\v x',\v x) = \bar{\Psi}_{s'}(\v x')\gamma^i\Psi_s(\v
x)$. The internal energy coming from the statistical interaction
$\braket{V_{\mathrm{st}}}$ is a linear function of the exciton order
parameter $\braket{\bar{\Psi}_s(\v x)\gamma^3\gamma^5\Psi_{s}(\v x')}$. As a
result, the mass has the same sign as that of the corresponding matrix
elements in $\mathcal{K}$ in order to lower the energy. To gain further insight,
one can isolate the interaction $V_{\mathrm{st}}$ in the exchange channel
(e.g., for the $R$-valley)
\begin{align}
  \label{eq:exchange_st}
  F_{\mathrm{st}} =&
  -\frac{\pi}{\mathcal{V}}\sum_{s,\v k,\v p}\mathcal{K}_{ss}\frac{\sin
    2\varphi_{Rs\v k_<}\cos 2\varphi_{Rs\v k_>}}{k_>}
  a^\dagger_{s\v k}a_{s\v k}a^\dagger_{s\v p}a_{s\v p}
\end{align}
where $\v k_>$ ($\v k_<$) is the bigger (smaller) of $\v k$ and $\v p$. To
lower the energy, i.e. to have a negative $F_{\mathrm{st}}$,
$\mathcal{K}_{ss}$ and $\varphi_{Rs\v k}$ must have the same sign and a
positive $\varphi$ implies a positive mass. The diagonal elements of
$\mathcal{K}^1$ have opposite signs, leading to opposite masses for the
spin-down and spin-up CDF bands. In contrast, the diagonal elements of
$\mathcal{K}^3$ have the same sign and thus produce mass gaps that are all
positive. For zeroth Landau level, the energy is given by the mass, e.g., if
the mass is negative, the energy of the zeroth Landau level will be
negative. Thus for different $\mathcal K$-matrices, the $\zeta LL$s have
different configurations as illustrated in Fig.~\ref{fig:landaulevels}. More
importantly, the transport property of Dirac fermions depends on the sign of
the mass since the direction of the Hall current is different for different
masses when the chemical potential lies in the mass gap. This property is
characterized by a topological number called the Chern number~\cite{TIReview}.

The exciton condensation is thus essential for classifying the different
states described by the three types of the ${\cal K}$ matrices. The
difference lies in the topology characterized by the Chern number of the CDF
bands, which is determined by the sign of the mass. Indeed, from the sign of
the mass gap, one can read out the Chern number of different states
described by the ${\cal K}$ matrices. For $\mathcal{K}^1$, the SU(4)
symmetry breaking leads to different signs in the mass gaps. The occupied
bands has a zero total Chern number and so the total filling factor is given
by $\nu=\widetilde\nu=\pm1/\widetilde\phi$. For the SU(4) symmetric case
described by ${\cal K}^3$, the identical mass gaps contribute to a
non-vanishing total Chern number $\pm2$, resulting in a quantum anomalous
Hall effect (QAHE) for the CDFs. Remarkably, this QAHE at $\widetilde\nu$
implies that the total filling factor $\nu=\pm(2 - 1/\widetilde\phi)$ for
the electrons, where the integer contribution to $\sigma_{xy}$ comes from
the QAHE of the exciton condensate. Similarly, $\mathcal{K}^2$ describes
states at filling factor $\nu = \pm(1 - 1/\widetilde \phi)$ since it leads
to CDF bands with Chern number $\pm 1$. In the rest of the paper, our focus
will be on the FQHE in the vicinity of charge neutrality observed in recent
experiments \cite{Du,Bolotin,Ghahari,Dean}.

We now calculate the dynamical mass for the symmetry breaking state
$\ket{N}$ described by ${\cal K}^1$ at $\nu = 1/2$ by minimizing the
variational energy $E_N = \braket{N|H|N}$ where the Hamiltonian includes
both the statistical interaction $V_{\rm st}$ and the Coulomb interaction
\begin{align}
  V_{\mathrm{c}}(\v x, \v x') = {g\over |\v x - \v x'|}, \quad g =
  {e^2\over 4\pi\varepsilon\hbar v_F}.
\end{align}
We focus on the spin-down bands and drop the spin indices for
simplicity. The variational equation for the $R$-valley (similar for the
$L$-valley) $\varphi_{R\v k}$ can be expressed as a set of self-consistent
equations for the quasiparticle dispersion $\epsilon_{R\v k} =
\sqrt{\alpha^2_{R\v k} + m^2_{R\v k}}$, where $m_{R\v k}$ is the mass gap,
$\alpha_{R\v k}$ is the renormalized dispersion, and $\sin 2\varphi_{R\v k}
= m_{R\v k}/\epsilon_{R\v k}$,
\begin{align}
  \label{eq:mass_normal}
  \alpha_{R\v k} = & k + g\frac{\pi}{\mathcal{V}}\sum_{\v p}v_1(k,p)
  \frac{\alpha_{R\v p}}{\epsilon_{R\v p}}d^R_{\v p}
  + \widetilde{\phi}\frac{2\pi}{\mathcal{V}}
  \sum_{p<k}\frac{1}{k}\frac{m_{R\v p}}{\epsilon_{R\v p}}d^R_{\v p},\nonumber \\
  m_{R\v k} =& g\frac{\pi}{\mathcal{V}} \sum_{\v p}v_0(k,p)\frac{m_{R\v
      p}}{\epsilon_{R\v p}}d^R_{\v p}
  + \widetilde{\phi}
  \frac{2\pi}{\mathcal{V}}\sum_{p > k}\frac{1}{p}
  \frac{\alpha_{R\v p}}{\epsilon_{R\v p}}d^R_{\v p}.
\end{align}
Here, $d^R_{\v k} = 1 - n^R_{\v k}$, $n^R_{\v k}$ is the occupation number
of the $R$-valley, and $v_\ell(k,p)$ is the coefficient of the angular
expansion of the Coulomb interaction $V_{\mathrm{c}}$ in the $\ell$-th
angular momentum channel, with the expansion given by
\begin{align}
  \label{eq:coulomb_expansion}
  \frac{1}{|\v k - \v p|} = \sum_{l = -\infty}^\infty
  v_{l}(k,p)e^{il(\theta_{\v k} - \theta_{\v p})}.
\end{align}

Since $d^R_{\v k}$ projects out the filled states, only the states above the
Fermi level contribute to the dynamical mass. Note that a natural
ultraviolet energy cutoff for Eqs.~\eqref{eq:mass_normal} is the energy
spacing between the $\zeta LL$ and the first $LL$, which is the largest
energy scale in the problem. Restoring the unit, the mass can be expressed
as $M_{R\v k} = m_{R\v k}\hbar k_F/v_F$ and scales with the external
magnetic field according to $M_{R\v k}\propto \sqrt{B}$. The magnitude of
$M_{R\v k}$ depends on the coupling constants. In particular, it depends on
$k_F\widetilde{\phi} \propto {\widetilde{\phi}}^{1/2}$. With increasing
$\widetilde{\phi}$, $M_{R\v k}$ will also increase. Therefore, for smaller
filling fractions, the dynamical mass will be larger. This agrees with the
fact that when filling fraction is small, there will be more states
available for exciton pairs, leading to larger masses. In the magnetic
catalysis theory~\cite{mc}, it is found that in the strong coupling regime
where Landau level mixing is relevant, the mass gap is proportional to the
Landau level spacing $\sqrt{2}\hbar v_F/l_B$, the same scaling behavior as
the mass gap obtained here. Since the CDFs are not confined to a single
Landau level, the current approach agrees well with the magnetic catalysis
theory in the strong coupling regime. It is worth emphasizing that, in
contrast to the nonrelativistic composite fermion theory where an
appropriate composite fermion mass remained elusive~\cite{HLR}, the mass of
the relativistic CDF theory naturally arises from the interactions through
exciton condensation.

\begin{figure}[htb]
  \centering%
  \includegraphics[width=0.45\textwidth]{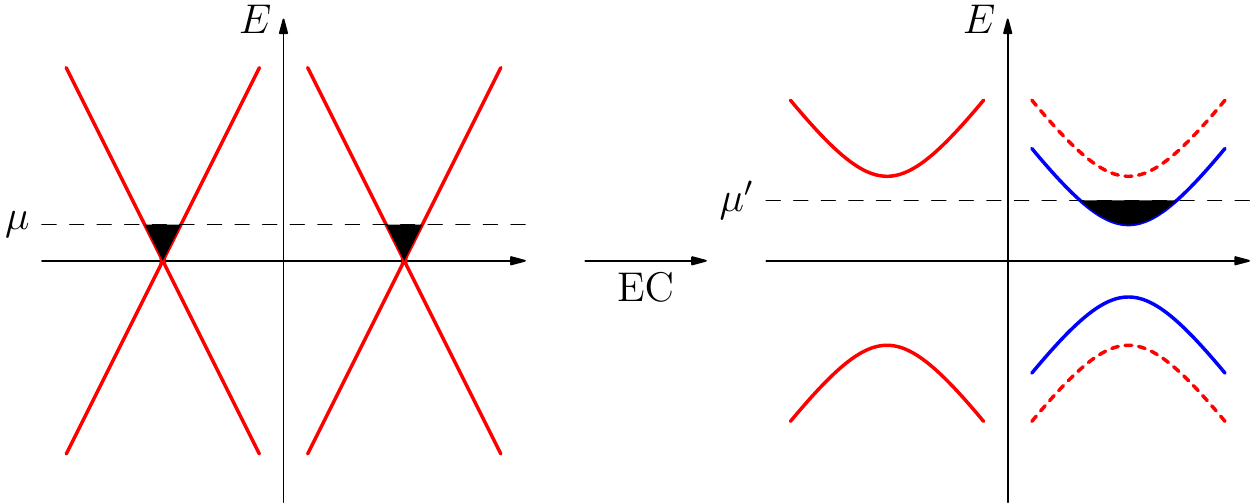}
  \caption{(color online) Schematic plot of the spin-down CDF
    bands before and after exciton condensation (EC) and exchange induced polarization.}
  \label{fig:schematic_bands}
\end{figure}

We found that, due to the exchange interaction, the valley polarized state
($k_F^R = \sqrt{2}k_F, k_F^L = 0$) has lower energy than the unpolarized
state ($k_F^R = k_F^L =k_F$). For example, for Coulomb strength $g=0.3$,
$\widetilde\phi = 2$ and a momentum cutoff $\Lambda = 2k_F$, the energy
density of the polarized state is approximately $-0.88$ while that of the
unpolarized state is $-0.85$. The solution for the complete dispersion of
the exciton mass $m_{R\v k}$ in the spin-valley polarized state is shown in
Fig.~\ref{fig:sc}a. Numerically, when $\v k$ is small, $M_{R\v k} \approx
0.55 E_{LL}$ where $E_{LL} = \sqrt{2}\hbar k_F/l_B$ is the Landau level
spacing between $n = 0$ and $n = 1$ Landau levels. However, due to its
momentum cutoff dependence, the exact magnitude of $M_{R\v k}$ must be
determined experimentally. The resulting CDF band structure is shown
schematically in Fig.~\ref{fig:schematic_bands}. The unoccupied spin-up
bands have a mass with the same magnitude but opposite sign as the
unoccupied spin-down band dictated by the diagonal elements of
$\mathcal{K}^1$ as discussed before. It should be stressed that the mass gap
of the empty CDF bands generated by the statistical interaction is large
enough so that the chemical potential lies inside the gap, making the
emergence of the spin-valley polarized state fully self-consistent with
$\mathcal{K}^1$.

In the case of $\nu = \pm 1/3$, these $\mathcal{K}$
matrices with $\widetilde{\phi} = 2$ does not lead to the complete
cancellation of the external magnetic field. Rather, the CDFs fill the Landau
levels of the residual magnetic field $B^* = B - 2|\rho|h/e$ at an integer
filling. These CDF Landau levels will develop mass gaps through the
statistical interaction and Coulomb interaction via magnetic catalyst at
integer filling factors~\cite{mc}. Under exchange splitting, only one CDF
Landau level will be completely filled and exhibits an integer QHE. This
corresponds to a FQHE state at $\nu= \pm 1/3$ of the electrons~\cite{Jain}
described by the single-component Laughlin state. In principle, other
states at $\nu = q/(2pq \pm 1)$ can be constructed following the composite
fermion approach~\cite{Toke3}. Within the current framework, however, we can
only study states within $\zeta LL$s due to the restriction that fictitious
magnetic field generated through flux attachment must partially or
completely cancel the external magnetic field. 

Before turning to the pairing instability, we comment on the relation
between magnetic catalysis and quantum Hall ferromagnetism. Recent
experiments~\cite{Dean, Young2012} showed that SU(4) symmetry of the $n =
\pm 1$ Landau levels is fully lifted. This seems to favor the quantum Hall
ferromagnetism theories since magnetic catalysis cannot explain the extra
plateaus at $\nu = \pm 3, \pm 5$. However, it is possible that the two
mechanisms (magnetic catalysis and quantum Hall ferromagnetism) play leading
roles in different regions and may even work in a collaborative way. For
example, it is plausible that magnetic catalysis may be the driving force in
the zeroth Landau levels while quantum Hall ferromagnetism is responsible
for Landau level splitting beyond the zeroth Landau level. Furthermore, it
was shown in Refs.~\cite{Gorbar2002, Semenoff} that the order parameters for
these two mechanisms can both be nonzero. 

\section{Paired Quantum Hall States}
\label{sec:ps}

\begin{figure}[htb]
  \centering%
  \includegraphics[width=0.48\textwidth]{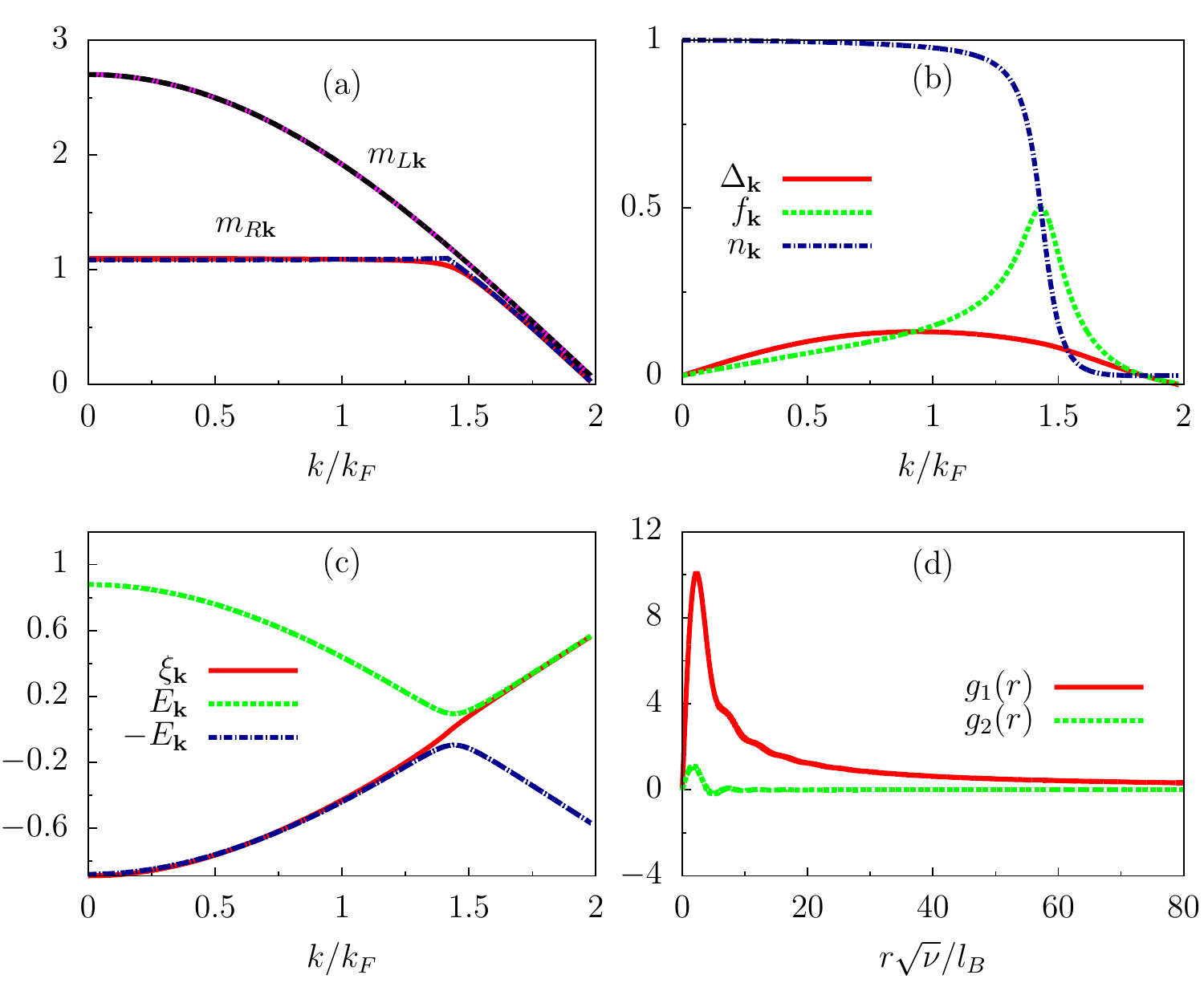}
  \caption{(color online) (a) The dynamical masses $m_{R\v k}$ and $m_{L\v k}$
    in the chiral $p$-wave paired state. The corresponding exciton masses in
    the normal state are very close and drawn in dashed lines. (b) Solution of
    the BdG equation for the pairing gap function $\varDelta_{\v k}$, the
    condensate amplitude $f_{\v k} = u^*_{\v k}v_{\v k}$, and the momentum
    distribution function $n_{\v k}$. (c) The CDF dispersion in the normal
    state and that of quasiparticle excitations in the paired state; (d)
    Pairing wave functions (unnormalized) in real space. All
    results are for $\widetilde{\phi} = 2$ and $g = 0.3$ with a momentum
    cutoff $\Lambda = 2k_F$. Due to valley splitting, the Fermi vector is
    $k^R_F = \sqrt{2}k_F$.}
  \label{fig:sc}
\end{figure}

We now show that, at even denominator filling fractions, the spin-valley
polarized composite massive Dirac fermion liquid has a pairing instability
where the quasiparticles on top of the exciton condensate form spin-triplet
pairs in the chiral $p$-wave channel. In terms of the $R$-valley, the
statistical pairing interaction is dominated by the $\ell=1$ angular
momentum channel and has the form,
\begin{align}
  \label{eq:pairing_st}
  P_{\mathrm{st}} =& -\frac{\pi}{\mathcal{V}}\sum_{s,\v k,\v
    p}\mathcal{K}_{ss} \frac{\sin 2\varphi_{Rs\v k_>}\cos
    2\varphi_{Rs\v
      k_<}}{k_>}\nonumber \\
  &\quad\times e^{i(\theta_{\v k} - \theta_{\v p})}a^\dagger_{s\v k}a^\dagger_{s-\v
    k}a_{s-\v p}a_{s\v p}.
\end{align}
It is remarkable that this pairing interaction is present only if there is
an exciton condensate, i.e., when $\varphi_{Rs\v k}\neq0$. Since the latter
requires Landau level mixing, this implies that Landau level mixing is
crucial for the pairing to occur. The variational wavefunction for the
paired state has the BCS form $\ket{\Omega} = \prod_{\v k}(u_{\v k} + v_{\v
  k}a^\dagger_{\v k}a^\dagger_{-\v k})\ket{0}$, with $|u_{\v k}|^2 + |v_{\v
  k}|^2 =1$, where $\vert 0\rangle$ is the exciton vacuum defined
in~\eqref{eq:exciton_vacuum}. Note that the variational wavefunction
contains both the exciton and pairing order parameters which must be
determined self-consistently by minimizing the ground state energy. The
self-consistent equations for the dynamical mass have the same form as in
\eqref{eq:mass_normal} with $n^R_{\v k} = |v_{\v k}|^2$, but unlike in the
normal state, exciton pairs exist even for $k < k^R_F$ due to pairing. The
variation of the energy with respect to $u^*_{\v k}$ and $v^*_{\v k}$ leads
to the familiar BdG equations (dropping the valley index),
\begin{align}
  \label{eq:bdg}
  E_{\v k}u_{\v k} &= \xi_{\v k}u_{\v k} + \varDelta^*_{\v k}v_{\v k},
  \nonumber \\
  E_{\v k}v_{\v k} &= -\xi_{\v k}v_{\v k} + \varDelta_{\v k}u_{\v k},
\end{align}
where $\xi_{\v k} = \epsilon_{\v k} - \beta_{\v k} - \mu$, $E_{\v k} =
\sqrt{\xi^2_{\v k} + |\varDelta_{\v k}|^2}$, and $\beta_{\v k} =
g\frac{\pi}{\mathcal{V}}\sum_{\v p}v_0(k,p)(1 + n^R_{\v p})$ comes
from the Coulomb exchange. The gap function $\varDelta_{\v k}$ is determined by the
gap equation
\begin{align*}
  \varDelta_{\v k} =& \frac{2\pi}{\mathcal{V}}\sum_{\v p}e^{i(\theta_{\v k}-\theta_{\v p})}u^*_{\v
    p}v_{\v p}\left[\widetilde{\phi}\frac{1}{k_>}\frac{m_{\v
        k_>}}{\epsilon_{\v k_>}}\frac{\alpha_{\v k_<}}{\epsilon_{\v k_<}}\right.\\
  -&\left. \frac{g}{2}\left(1 + \frac{m_{\v k}m_{\v p}}{\epsilon_{\v
          k}\epsilon_{\v p}}\right)v_1(k,p) - \frac{g}{2}\frac{\alpha_{\v
        k}\alpha_{\v p}}{\epsilon_{\v k}\epsilon_{\v p}}v_0(k,p)\right].
\end{align*}
Fig.~\ref{fig:sc} displays the numerical solution of $E_{\v k}$ and $\varDelta_{\v k}$ at
$\widetilde \phi=2$ and $g=0.3$, where the
ground state is indeed a chiral $p+ip$ paired state. Because the Coulomb
interaction is pair-breaking, the pairing gap
$\Delta_{\mathrm{sc}}=2{\rm min}(E_{\v k})$ reduces with increasing $g$ and
vanishes at a critical value $g_c \approx 0.53$ and $1.28$ for
$\widetilde{\phi} = 2$ and $4$ respectively. For $g>g_c$, the massive Dirac
fermions form a stable Fermi liquid state.

To gain further insights into the paired state, we study the two-particle
pairing wave function by projecting the BCS state to real space: $g(\v x_1 -
\v x_2) = \braket{0|\Psi(\v x_1)\Psi(\v x_2)|\Omega}$. Because of the spinor
structure, the upper and lower components are obtained separately,
\begin{align}
  \label{eq:pwf}
  g_1(\v x) &= \frac{1}{\mathcal{V}}\sum_{\v k}\left(\cos\varphi_{R\v k} +
    \sin \varphi_{R\v k}\right)^2e^{-2i\theta_{\v k}}e^{i\v k\cdot\v x}
  g_{\v k},\nonumber \\
  g_2(\v x) &= \frac{1}{\mathcal{V}}\sum_{\v k}\left(\cos\varphi_{R\v k} -
    \sin \varphi_{R\v k}\right)^2e^{i\v k\cdot\v x}
  g_{\v k},
\end{align}
where $g_{\v k} = v_{\v k}/u_{\v k}$. In the long wavelength limit, $g_{\v
  k}\approx e^{i\theta_{\v k}}/k$ and $m_{R\v k}\approx m_0$ is essentially
a constant (c.f. Fig.~\ref{fig:sc}). We have, to quadratic order,
$(\cos\varphi_{R\v k} + \sin \varphi_{R\v k})^2 \approx 2 -
{v_F^\prime}^2k^2/2m^2_0$ and $(\cos\varphi_{R\v k} - \sin \varphi_{R\v
  k})^2 \approx {v_F^\prime}^2 k^2/2m^2_0$ where $v_F^\prime$ is the
renormalized Fermi velocity. Thus, the large distance behaviors of the
pairing wave function are $g_1(r) \propto 1/z + c\sqrt{2/\pi}\cos(\kappa r -
\pi/4)/z\sqrt{r}$ and $g_2(r) \propto - c\sqrt{2/\pi}\cos(\kappa r -
\pi/4)/\bar{z}\sqrt{r}$, where $c$ and $\kappa$ are numerical constants. The
oscillatory terms in the above equations are due to the $k^2$ terms in the
expansions, originating from the $k$ dependence of the mass, which should be
distinguished from the oscillatory wave function in
Ref.~\cite{nonunitary}. The many-body real space wave function for CDFs can
be obtained by~\cite{RG}
\begin{align*}
&  \Psi_{\mathrm{CDF}}(\v x_1, \v x_2, \ldots, \v x_N) = \braket{0|\Psi(\v x_1)\Psi(\v
  x_2)\ldots\Psi(\v x_N)|\Omega} \\
&= \frac{1}{2^{\frac{N}{2}}\left(\frac{N}{2}\right)!}\sum_P \mathrm{sgn}(P)
\begin{pmatrix}
\prod_{i = 1}^{N/2}g_1(\v x_{P(2i - 1)} - \v x_{P(2i)})\\
 \prod_{i = 1}^{N/2}g_2(\v x_{P(2i - 1)} - \v x_{P(2i)})
\end{pmatrix}\\
&\approx \frac{1}{2^{\frac{N}{2}}\left(\frac{N}{2}\right)!}\sum_P \mathrm{sgn}(P)
\begin{pmatrix}
\prod_{i = 1}^{N/2} 1/(z_{P(2i - 1)} - z_{P(2i)})\\
0
\end{pmatrix}.
\end{align*}
where $N$ is an even number representing the number of CDFs. In the last
line of the above equation, quadratic and higher order terms have been
dropped. Note that except for the subleading oscillatory contributions that
decay faster at large distances, the pairing wave function resides
predominantly on one component of the spinor (i.e. on one of the
sublattices) and has the form of the Moore-Read Pfaffian state in agreement
with the numerical results shown in Fig.~\ref{fig:sc}. It is remarkable that
although the large holomorphic part on the upper component is indeed
dominated by contributions from the $\zeta LL$ subspace, the wavefunction of
the nonabelian ground state does not entirely lie in the $\zeta LL$ since
the nonholomorphic, oscillatory contributions, although small, enter both
the upper and the lower components of the wavefunction and can be attributed
to the effects of Landau level mixing.

Similar to Ref.~\cite{HLR, Read95}, from the transformation in
Eq.~\ref{eq:unitary_transformation}, the wave function for Dirac fermions is
related to that of CDF by
\begin{align*}
  \Psi_{\mathrm{MF}}(\v x_1,\ldots,\v x_N) = \prod_{i<j}\frac{(z_i -
    z_j)^{\widetilde{\phi}}}{|z_i -
    z_j|^{\widetilde{\phi}}}\Psi_{\mathrm{CDF}}(\v x_1, \v x_2, \ldots,\v x_N).
\end{align*}
The full Jastrow factor $\prod_{i<j}(z_i - z_j)^{\widetilde{\phi}}$ may be
recovered by taken into account of fluctuations beyond mean field
level~\cite{Zhang92}.

\section{Conclusion}
\label{sec:conclusion}

In this paper, we presented a composite relativistic fermion theory for the
FQHE in graphene. We showed that the ground state has spontaneous
spin-valley polarization, leading to a single-component Laughlin-like state
at $\nu=1/3$ and a chiral $p$-wave pairing state of the Moore-Read Pfaffian
at $\nu=1/2$. A crucial prediction of the present theory is the dynamical
mass generation for the Dirac fermions through the formation of an exciton
condensate in the zeroth Landau level. It originates from Landau level
mixing and facilitates the spin-valley polarization and the pairing
interaction for the paired states at even dominator filling fractions. This
CDF mass, scaling with $\sqrt{B}$, should be detectable by scanning
tunneling microscopy in high magnetic field. It would also be desirable to
see if and how the inclusion of Landau level mixing affects the results of
the numerical diagonalization studies~\cite{Toke2,Wojs}. In a recent
experiment~\cite{Ghahari} on suspended graphene, a plateau like feature near
$\nu = 1/2$ is observed in some but not all samples. Future experiments are
desirable to explore whether a true single-component quantum Hall state
emerges at $\nu=1/2$ that would realize nonabelian statistics in graphene.

We end this paper by a few comments on some experimental results. For
definiteness, we assumed near charge neutral state, the spin is polarized. A
recent experiment~\cite{Young2012} shows that the $\nu = 0$ state is spin
unpolarized. This suggests that the valley anisotropy is the dominant force
behind the symmetry breaking near charge neutrality. This effect can be
easily incorporated in the current framework by including a valley
anisotropy term. In Ref.~\cite{Dean}, it is found that $\nu = 5/3$ state is
missing. This may be attributed to the different environment this state lies
in. For $\nu = 1/3$ state, the associated $\mathcal{K}$-matrix breaks the
SU(4) symmetry. In contrast, the $\mathcal{K}$-matrix for $\nu = 5/3$
preserves the symmetry. Due to the SU(4) symmetry, the associated Goldstone
modes in $\nu = 5/3$ may destabilize the state. A discussion of this issue
can be found in Ref.~\cite{Toke3,Abanin2013}. A more surprising result in
Ref.~\cite{Feldman1} is that the energy gaps are linear in $B$. Given that
the energy scales in the problem are Coulomb interaction energy and Landau
level spacing both scaling as $B^{1/2}$, any theory based on these energy
scales should produce an energy gap that scales with $B^{1/2}$. Thus, this
result presents a great puzzle. Further experimental and theoretical studies
are needed to resolve it.

This work is supported in part by DOE grant DE-FG02-99ER45747,
the NNSF of China, and the 973 program of MOST of China.

\end{document}